\documentclass[%
 reprint,
 amsmath,amssymb,
 aps,
]{revtex4-2}

\usepackage{graphicx}
\usepackage{dcolumn}
\usepackage{bm}
\newcommand{\braket}[1]{\langle #1 \rangle}
\begin{document}

\preprint{}

\title{Primordial black hole mass growth from neutrinos during the radiation era}

\author{Ma\"el Gonin}
 \email{mael.gonin@dzastro.de}
 \affiliation{TU Dresden, Institute of Nuclear and Particle Physics (IKTP), 01062 Dresden, Germany}
 \affiliation{Deutsches Zentrum f\"ur Astrophysik (DZA), Postplatz 1, 02826 G\"orlitz, Germany}

\date{\today}

\begin{abstract}
We present a new picture of primordial black hole mass evolution through neutrino 
absorption. Using a semi-classical approach and a detailed look at the kinetics of the 
early plasma, we revisit the thermal absorption of radiation by a population of primordial 
black holes ranging from $10^{-3}$ to $10^9\,M_\odot$, embedded in a thermal bath. We 
find significant mass growth for intermediate-mass and supermassive PBHs; this growth 
shifts the predicted $e^+e^-$ peak in the mass spectrum arising from the thermal history. 
Depending on the value of the collapse fraction $\gamma$, an additional peak in the 
intermediate-mass range may become significant. Furthermore, since PBHs grow by accreting 
energy from the thermal bath, the fraction of dark matter in PBHs, $f_{\rm PBH}$, also 
changes. These results revise the standard picture of PBH mass evolution and have 
implications for observational constraints on PBHs as dark matter candidates.
\end{abstract}

\maketitle


\section{Introduction}
\label{sec:intro}
Black Holes (BH) studies and modern cosmology both stem from the general relativistic description of gravity. Primordial Black Holes (PBHs) provide a direct connection between these two fundamental fields of physics. PBHs as Dark Matter (DM) candidates have been considered for a long time in the literature \cite{Carr:1974nx,Carr:1975qj,Carr:2020gox,Byrnes:2025tji,Carr:2026hot}. A plethora of phenomena could lead to gravitational collapse in the early universe, the most studied being probably the collapse from inflation-generated overdensities \cite{LISACosmologyWorkingGroup:2023njw}.
Gravitational wave detections since 2015 and the subsequent catalog \cite{LIGOScientific:2025slb,LIGOScientific:2026wfs} have revitalized interest in PBHs \cite{Carr:2019kxo}. Such populations of BHs have already been suggested to explain GW observations \cite{Bird:2016dcv,Sasaki:2016jop,Clesse:2016vqa}. Many observations could constrain PBH abundance over a broad mass range \cite{LISACosmologyWorkingGroup:2023njw,Carr:2026hot}. These often rely on the assumption of a monochromatic PBH mass spectrum. While convenient for estimations, the monochromatic assumption is hardly justified given the theoretical uncertainties in inflation scenarios. PBH clustering could also undermine the validity of constraint estimates \cite{Carr:2026hot}.

Moreover, established physical scenarios such as Choptuik's law \cite{Choptuik:1992jv,Evans:1994pj,Niemeyer:1999ak} on the criticality of gravitational collapse naturally break down the monochromatic assumption by broadening the peak. The thermal history of the Universe impacts the cosmic Equation of State (EoS) \cite{Laine:2015kra,Borsanyi:2016ksw,2018JCAP...08..041B,Gonin:2025uvc,Ferreira:2025zeu,Formaggio:2025nde,Gonin:2026xhe} and creates peaks in the PBH mass distribution at specific masses \cite{2018JCAP...08..041B,Carr:2019kxo,Bodeker:2020stj,Gonin:2025uvc,Gonin:2026xhe}. We can therefore predict characteristic PBH masses based on ground-based experiments and simulations of fundamental interactions. This entanglement between primordial plasma physics and the mass spectrum makes PBHs probes of the early Universe \cite{Bodeker:2020stj,Khlopov:2024nqp}.
The active debate over the validity of constraints and what conclusions one can draw from them requires an accurate description of the PBH history from its very beginning to today.

The idea of PBHs is about the possibility of forming BHs in the primordial plasma \cite{Zeldovich:1967lct,Carr:1974nx}, when the Universe was radiation dominated. In 1967, Zel'dovich and Novikov estimated that such objects would absorbed the surrounding radiation, leading to a catastrophic mass growth in the radiation era \cite{Zeldovich:1967lct}, while in 1974 Carr and Hawking discarded Zel'dovich's argument \cite{Carr:1974nx}. They invoked that the mean free path of radiation $\lambda^{\rm rad}_{\rm mfp}$, would be much smaller than the PBH Schwarzschild radius $r_S(M_{\rm PBH})$ and than the cosmic horizon; hence, PBHs would accrete in a hydrodynamical regime that is not efficient enough for significant mass growth.

Since this foundational work, the general view on PBHs is that they are not able to gain mass in the radiation era \cite{Custodio:1998pv,Custodio:2002gj}. We re-examine this view, arguing that a closer look at primordial plasma kinetics actually allows for neutrino radiation absorption. We do not discard the Carr and Hawking argument, but rather state that the condition on the radiation mean free path does not hold at all times. The mechanism presented in this letter does not rely on hydrodynamical calculations of a radiation fluid, but rather treats BHs as black body radiation absorbers. The present work is based on the recent study Ref.~\cite{Haque:2026vvp}.

In Section \ref{sec:PBH_spectrum}, we present the PBH mass spectrum and its features. In Section \ref{sec:NuAbsorption}, we review the theory of PBH absorption and how neutrinos could be absorbed. In Section \ref{sec:results}, we present the results on the extended PBH mass spectrum.
Throughout this study we use natural units : $\hbar=c=k_B=1$.

\section{The extended PBH mass spectrum from thermal history
}
\label{sec:PBH_spectrum}

PBHs can form from a variety of processes; see recent reviews \cite{LISACosmologyWorkingGroup:2023njw,Byrnes:2025tji,Carr:2026hot}. In this section, we focus on describing the PBH mass spectrum from the thermal history \cite{2018JCAP...08..041B,Carr:2019kxo,Bodeker:2020stj}; see also Husdal ~\cite{Husdal:2016haj} for detailed descriptions of the EoS. Throughout this study, we use the code \texttt{CosmicEoS}; see~\cite{Gonin:2025uvc,
Gonin:2026xhe} and references therein.
Formation from the collapse of inflationary fluctuations is often invoked, as it does not require new physical phenomena. Instead, it builds on the theoretical uncertainties regarding inflation scenarios. Many of these scenarios can leave a tail at small scales \cite{Franciolini:2022tfm}. This is our setup: the Universe is filled with overdense regions of various sizes characterized by the energy density contrast \cite{Carr:1974nx}:

\begin{equation}
    \delta=\frac{\rho-\rho_B}{\rho_B}
\end{equation}
where $\rho$ is the energy density in the overdense region and $\rho_B$ is the background energy density.
 A PBH is formed if $\delta\geqslant \delta_c$ at the moment when the particle horizon crosses the radius \cite{Carr:1975qj}. $\delta_c$ is the critical density threshold, a function of the plasma EoS $w=P/\rho$ \cite{Escriva:2022bwe,Franciolini:2022tfm,Musco:2023dak}. 
The softening of $w$ during cosmic phase transitions cause $\delta_c$ to decrease; therefore, cosmic phase transitions increase PBH production. Peaks appear in the mass distribution, making PBHs unique probes of the plasma EoS \cite{2018JCAP...08..041B,Carr:2019kxo,Bodeker:2020stj,Escriva:2022bwe,Franciolini:2022tfm,Musco:2023dak,Gonin:2025uvc,Gonin:2026xhe}.

Since gravitational collapse is a critical phenomenon, the value of the difference $\delta-\delta_c$ impacts the PBH mass \cite{Musco:2012au,Musco:2023dak}. 
Due to pressure gradients and criticality, PBHs are not born with exactly the horizon mass at crossing time, i.e., the overdensity mass. To account for this discrepancy, the collapse fraction $\gamma$ is often invoked as:
\begin{equation}\label{eq:M_PBH}
    M_{\rm PBH}=\gamma M_H
\end{equation}
where $M_H$ is the mass enclosed in the cosmic horizon at crossing time:
\begin{equation}
    M_H(T)=\frac{4}{3}\pi\rho_B(T)R_H^3
\end{equation}

Refs.~\cite{Musco:2012au,Escriva:2022bwe,Musco:2023dak} showed that the collapse fraction depends on the criticality of the collapse, the PBH mass, and the shape of the fluctuation. $\gamma$ is a function of these parameters. The original spectrum in this study was computed with $\gamma=1$.
Moreover, for spatially correlated peaks, more matter can fall into the PBH \cite{Choi:2025eqn}.

For simplicity, $\gamma$ is treated as a free parameter in this study. 
We assume a Gaussian spectrum of fluctuations and rely on Press-Schechter statistics to evaluate PBH abundance; the procedure has been detailed in Refs.~\cite{2018JCAP...08..041B,Carr:2019kxo,Bodeker:2020stj,Escriva:2022bwe,Franciolini:2022tfm,Musco:2023dak,Gonin:2025uvc,Gonin:2026xhe}. We quickly summarize the equations in Appendix \ref{app:pbh_spectrum}. The amplitude of the fluctuation spectrum, $A$ is a free parameter that is not restricted by observations on the scales considered. It is therefore the parameter used to normalize the fraction of dark matter in PBHs $f_{\rm PBH}=\Omega_{PBH}/\Omega_{DM}$.

The PBH spectra presented in this study should be viewed as a toy model. As mentioned in the section, many aspects of PBH collapse are ignored in the present study. Since we present a new mechanism, we want to limit the number of parameters and therefore stick to the toy model of an extended PBH mass spectrum.

\section{Neutrinos Absorption}
\label{sec:NuAbsorption} 

In this section, we present the procedure used to evaluate neutrino absorption by a BH 
embedded in the primordial plasma. Since we aim to revise the standard picture of PBH 
evolution in the radiation era, we begin this section with a brief summary of the 
discussion. In Ref.~\cite{Zeldovich:1967lct}, Zel'dovich and Novikov treat the primordial 
plasma as a non-interacting ($\lambda^{\rm rad}_{\rm mfp} > r_S$) and relativistic gas 
surrounding a compact object~\footnote{We refer to a plasma fulfilling this set of 
conditions as a `thermal bath', following Refs.~\cite{Barrau:2022bfg,Barrau:2025psm,
Haque:2026vvp}.}
, giving it a cross section $\sigma \propto r_S^2$~\cite{1966SvPhU...8..522Z,
Zeldovich:1967lct} and concluding that accretion of radiation could be catastrophically 
large. Carr and Hawking in Ref.~\cite{Carr:1974nx} argued that the condition 
$\lambda^{\rm rad}_{\rm mfp} > r_S$ is hardly fulfilled due to the very small value of 
$\lambda^{\rm rad}_{\rm mfp}$. They relied on Bondi accretion~\cite{1952MNRAS.112..195B} 
to argue that a PBH cannot sustain growth comparable to the cosmological horizon, the 
accretion being limited to the hydrodynamic (Bondi) rate. Later, in the late 1990s and 
early 2000s, Custodio and Horvath published a series of papers~\cite{Custodio:1998pv,
Custodio:2002gj} on the evolution of PBHs in the radiation era, taking into account 
Hawking radiation. While never mentioning the condition on $\lambda^{\rm rad}_{\rm mfp}$, 
they conclude that mass growth is negligible; however, their radiation density (Eq.~6 
of~\cite{Custodio:2002gj}) corresponds to a relativistic degrees of freedom of 
$g_\rho \sim 1$, an order of magnitude below the Standard Model value at MeV temperatures, and enters their maximal gain quadratically.

Much more recently, Ref.~\cite{Das:2025vts,Haque:2026vvp} independently recovers the same evolution 
equation as Refs.~\cite{Custodio:1998pv,Custodio:2002gj} for PBH absorption. Ref.~\cite{Haque:2026vvp} builds a model for reheating and DM production from evaporating PBHs, while Ref.~\cite{Das:2025vts} examines general-relativistic accretion corrections. They acknowledge the 
condition on $\lambda^{\rm rad}_{\rm mfp}$, argue that it is eventually fulfilled for 
these small PBHs, and conclude that absorption allows evaporating PBHs to extend their 
lifetime. Interestingly, another preprint, Ref.~\cite{Kallifatides:2026sik}, released the 
same week, presents similar calculations of PBH mass growth. We acknowledge this study but 
do not make use of their method in the present letter. See also Ref.~\cite{Chatterjee:2025wnt} for a study
of the impact of the thermal bath on the Hawking emission.

Note that Hawking radiation may not be the only spontaneous particle production process 
associated with BHs; superradiance could also play a role. For hot (small) BHs, 
superradiance could limit the ability of a BH to absorb a fermionic flux~\cite{Dai:2023zcj}.

Moreover, Barrau et al. in Ref.~\cite{Barrau:2022bfg} discussed `catastrophic' mass 
growth (also referred to as `runaway absorption' in the present study) of a BH in a thermal bath, i.e.\ 
$M \rightarrow \infty$ for $t < \infty$. In Ref.~\cite{Barrau:2025psm} they propose a 
toy model to regularize the metric.

Our approach and the mass growth equations are based on the recent work in Ref.~\cite{Haque:2026vvp}. We explicitly state when different assumptions are being made. Contrary to the previous studies we take a closer look at the condition on $\lambda^{\rm rad}_{\rm mfp}$, using our state-of-the-art code \texttt{CosmicEoS} to model the thermodynamics of the primordial plasma we derive $\lambda^{\nu}_{\rm mfp}$ and find that mass growth from neutrino radiation is not only possible but can be significant.

A BH is said to be embedded in a radiation bath and can absorb the surrounding black body radiation if $T_{\rm BH}<T_{\rm rad}$ and $\lambda^{\rm rad}_{\rm mfp}>r_S(M_{\rm PBH})$ \cite{Zeldovich:1967lct,Haque:2026vvp}. The BH temperature is given by \cite{Hawking:1975vcx,Page:1976df}:

\begin{equation}
    T_{\rm BH}=M_{Pl}^2/M_{\rm BH}
\end{equation}
With $M_{Pl}\equiv M_P/\sqrt{8\pi}=2.18\times10^{-39}M_\odot=2.44\times10^{21} \rm~MeV$ the reduced Planck mass (whith $M_P=\sqrt{\hbar c/G}$ the Planck mass), it is clear that for the mass range considered, $T_{\rm BH}$ is extremely small, and we have $T_{\rm BH}<T_{\rm rad}$ at all time.

\begin{figure}
    \centering
    \includegraphics[width=\linewidth]{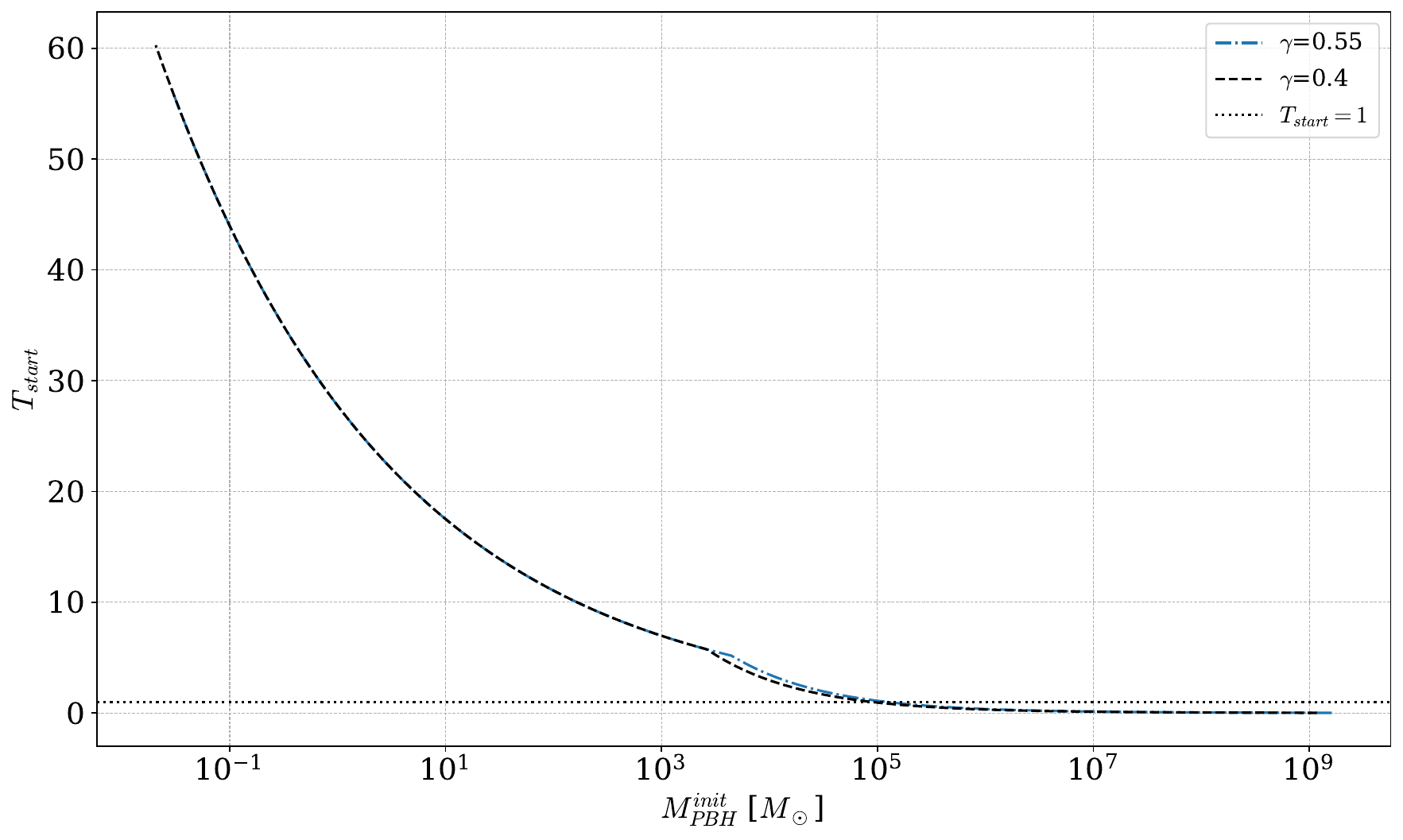}
    \caption{$T_{\rm start}(r_S)$ across the PBH mass mass range considered for various values of $\gamma$ from Eq.~\eqref{eq:M_PBH} and Eq.~\eqref{eq:gamma_eff}. }
    \label{fig:T_start}
\end{figure}

The Schwarzschild radius is given by: 
\begin{equation}
    r_S=2GM_{\rm BH}=\frac{M_{\rm BH}}{4\pi M_{Pl}^2}
\end{equation}
While the neutrino mean free path \cite{Dolgov:2002wy}:
\begin{equation}\label{eq:nuMFP}
    \lambda^{\nu}_{\rm mfp}=\frac{1}{n^{\rm tot}_\nu \braket{\sigma_{\rm weak}}}
\end{equation}
with $n^{\rm tot}_\nu=n_\nu+n_{\bar{\nu}}$ easily comes from Fermi-Dirac statistics for vanishing masses and $\braket{\sigma_{\rm weak}}=G_F^2  T^2 / \pi $ \cite{Kolb:1990vq}, the cross section of the weak interaction. We define a temperature $T_{\rm start}(r_S)$ corresponding to the start of neutrino absorption, i.e., 
$\lambda^{\nu}_{\rm mfp}(T_{\rm start})\geqslant r_S(M_{\rm PBH})$.
$T_{\rm start}$ for our PBH mass range is shown in Fig.~\ref{fig:T_start}. The inflection 
point corresponds to the moment at which the PBH fulfils $r_S(M_{\rm PBH}^{\rm init}) 
\lesssim \lambda^{\nu}_{\rm mfp}$ at formation, i.e.\ $T_{\rm start} = T_{\rm init}$.

We introduce the idea of a transition between the non-absorbing and absorbing regimes. A naive picture would be a step function when $\lambda^{\nu}_{\rm mfp}\geqslant r_S$ is true, but in the absence of certainty on this transition, we also model the smooth transition through a weight function:

\begin{equation}\label{eq:weight}
    W=e^{-\kappa r_S/\lambda^\nu_{\rm mfp}}
\end{equation}
where $\kappa=[0.5; 1;2]$ is the path length through the plasma in units of mean free path that allows a neutrino to be absorbed. Such a weight function captures the fact that a PBH might absorb for $r_S\lesssim \lambda^\nu_{\rm mfp}$, and that for $r_S\sim\lambda^\nu_{\rm mfp}$ the absorption is not fully efficient.


Once the PBH is effectively embedded in a thermal bath, it acquire a cross section. In this study we use the geometric optic limit allowed by $\omega \gg 1/r_S$ giving $\sigma \propto r_S^2$ \cite{1966SvPhU...8..522Z,Zeldovich:1967lct} (see Appendix \ref{app:MassGrowth}). Detailed descriptions of the BH cross section across frequency regimes were given in Refs.~\cite{Unruh:1976fm,Sanchez:1977si}. The convergence to the geometric-optics value has been established for spin 0 \cite{Sanchez:1977si}, spin 1/2 \cite{Unruh:1976fm,Doran:2005vm,Dolan:2006vj}, and spin 1 \cite{Crispino:2007qw}; hence the simple expression used throughout this study stands on solid theoretical ground.

We solve the following ODE to trace the mass evolution:
\begin{equation}\label{eq:ODE}
    \frac{dR}{dx}=\frac{12\pi \gamma_{\rm eff} \delta_{hf}}{\alpha x^3}R^3 W
\end{equation}

where $R=M_{\rm PBH}^{\rm abs}(x)/M_{\rm PBH}^{\rm rad}$, $x=a/a_{\rm rad}$, $\alpha=g_\rho^{\rm tot} \pi^2/30$ , $\delta_{hf}=\frac{9\pi}{640}g_\rho^\nu$ (see Appendix \ref{app:MassGrowth}), $W$ is defined by Eq.~\eqref{eq:weight} and we introduce the effective collapse fraction, $\gamma_{\rm eff}$ as: 
\begin{equation}\label{eq:gamma_eff}
    \gamma_{\rm eff}= \gamma(1+\delta_c(w(T)))
\end{equation}

With $\delta_c(w(T))$ from \cite{Musco:2012au}, see \cite{Bodeker:2020stj,Gonin:2025uvc} for the evolution along the temperature axis. In Eq.~\eqref{eq:gamma_eff}, we assume that all fluctuations forming a PBH have exactly $\delta=\delta_c$; in reality, $\delta\geqslant\delta_c$ could be realized. However, accounting for super-critical fluctuations would require the inclusion of the criticality of PBH formation. It is outside the scope of the current study. We discuss the choice of $\gamma_{\rm eff}$ in Appendix \ref{app:MassGrowth}.

Contrary to Ref.~\cite{Haque:2026vvp}, the temperature at formation $T_{\rm init}$ and $T_{\rm start}$ do not necessarily coincide. In Eq.~\eqref{eq:ODE}, $a_{\rm rad}$ corresponds to the formation time, but we solve the ODE over $[T_{\rm start};~T_{\rm eq}]$ where $T_{\rm eq}=0.8 \rm ~eV$ is the temperature at matter-radiation equality.

Since the ODE $dR/dt\propto x^{-3}=s(T)/s_{\rm rad}$, where $s$ is the entropy density, the absorption rate drops with time.
As $\gamma$ is a relevant parameter for mass growth, it is possible to constrain its value to avoid runaway absorption.
We define the diverging values applicable for $\frac{g_\rho^{\rm tot}(T)}{g_\rho^{\nu}(T)}=\rm const$ as (see Appendix \ref{app:MassGrowth} for the derivation and validity of the expression):
\begin{equation}\label{eq:gamma_div}
    \gamma_{\rm div}(T)=\frac{32}{81}\frac{g_\rho^{\rm tot}(T)}{g_\rho^{\nu}(T)}\frac{1}{1+\delta_c(T)}
\end{equation}

Ref.~\cite{Haque:2026vvp} puts a bound on $\gamma<32/81$, but those depend on the energy content of the plasma; we cannot apply their bounds to neutrinos absorption \footnote{
Interestingly, the critical value 32/81 already appears in Ref.~\cite{Zeldovich:1967lct} [Eqs.~(2)–(3) and surrounding discussion], predating its modern derivation \cite{Haque:2026vvp}.
}. Ref.~\cite{Kallifatides:2026sik} also worked on bounding $\gamma$ over a broader mass range. While these two studies might look analogous at first sight, they do not invoke the same mechanism. Ref.~\cite{Haque:2026vvp} is about absorption based on geometrical cross section, while Ref.~\cite{Kallifatides:2026sik} is based on Stefan-Boltzmann equilibrium between the BH and the thermal bath.

The updated PBH mass spectrum is defined by:
\begin{equation}\label{eq:abs_spectrum}
    \frac{df_{\rm PBH}^{\rm abs}}{dlnM} = R(M_{\rm PBH})\frac{df_{\rm PBH}^{\rm rad}}{dlnM}
\end{equation}


From Eq.~\eqref{eq:ODE}, $R$ depends on $\delta_{hf}/\alpha$, which is $\propto \rho_\nu/\rho_{total}$. This brings a dependency on the primordial asymmetries; when considering lepton asymmetry, neutrinos can take a large portion of the energy density \cite{Gonin:2026xhe}, hence more radiation is available and larger mass growth results. Although once again, this effect competes with the fact that the total number of neutrinos $n_{\rm tot}$ increases with lepton asymmetry, which makes $\lambda_{\rm mfp}^{\nu}$ smaller from Eq.\eqref{eq:nuMFP}. We defer such work to future studies \footnote{Additionally, in presence of lepton asymmetry the neutrinos flavor oscillate at the time of decoupling \cite{Froustey:2020mcq,Froustey:2024mgf}. The mechanism and the associated entropy release requires a careful treatment.}.

Having defined the equations and their validity domain, we can now compute the resulting mass growth.

\section{The absorbing PBH mass spectrum}\label{sec:results}

\begin{figure}
    \centering
    \includegraphics[width=1\linewidth]{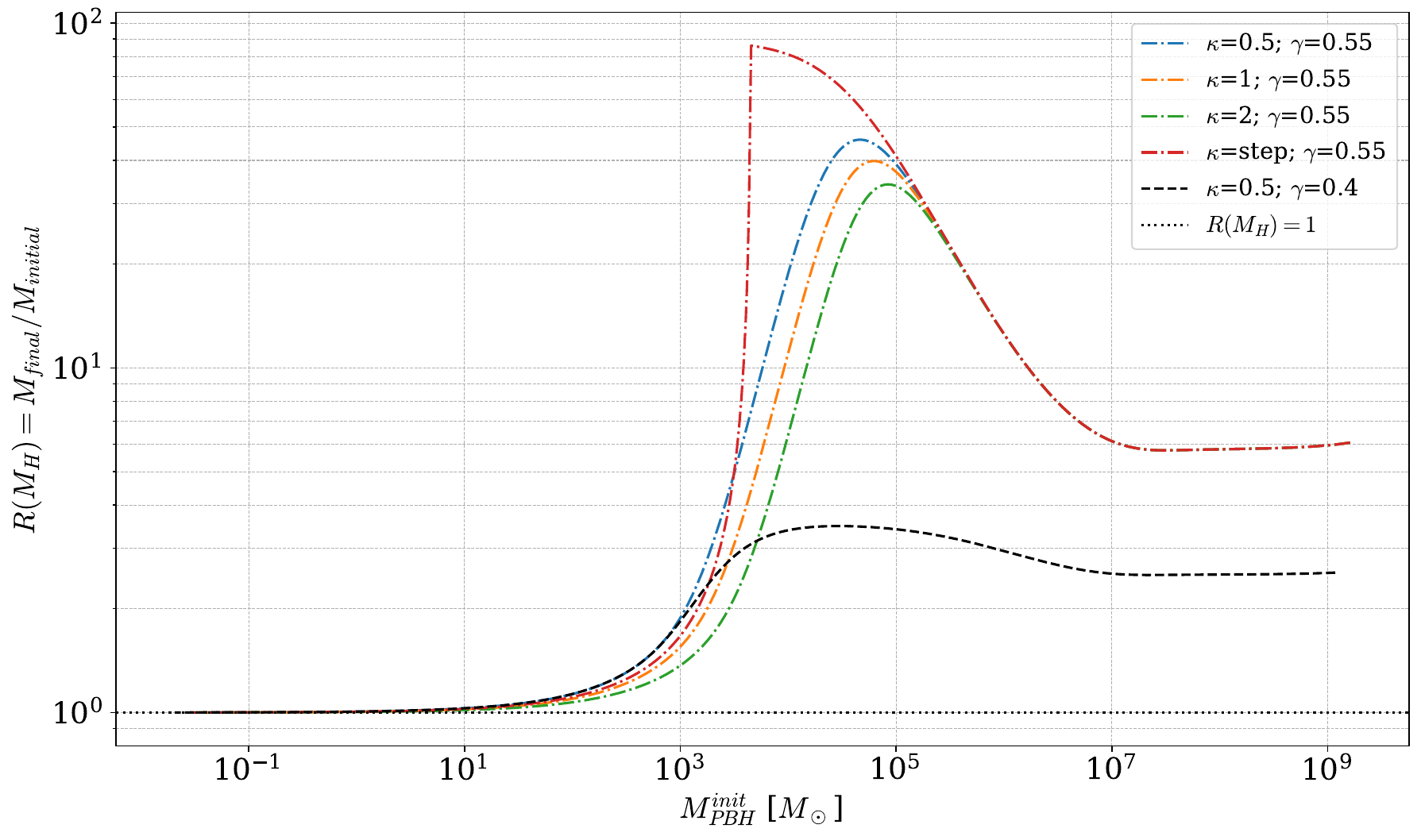}
    \caption{Mass growth $R$ from the ODE~\eqref{eq:ODE} in the PBH mass range considered with $\gamma=0.55$}
    \label{fig:Mass_growth}
\end{figure}

In this section, we present the results on mass growth from neutrino absorption, the subsequent updated PBH mass spectra, and the fraction of dark matter in PBHs after absorption, $f_{\rm PBH}^{\rm abs}$.
We plot the mass growth $R$ in Fig.~\ref{fig:Mass_growth} for various values of $\kappa$. Only PBHs with $M\geqslant10^3M_\odot$ significantly absorb neutrinos. The reason is that as $M_{\rm PBH}$ increases, $T_{\rm init}$ gets closer to $T_{\rm start}$, with the step function in red jumping when $T_{\rm start}=T_{\rm init}$.
One feature of the plot is also the slow rise of $R$ prior to the jump, meaning that smaller PBHs also absorb. However, since the difference between formation and absorbing times increase with decreasing PBH mass, it means that $x$ at the onset of absorption is very small; from Eq.~\eqref{eq:ODE}, $R$ depends sensibly on $x$. One of the results from \cite{Haque:2026vvp} is that the bulk of the mass growth happens in the few moments after PBH formation; there is a sweet spot for absorption to occur when $a\sim a_{\rm rad}$, hence the factor $x\lesssim1$  in Eq.~\eqref{eq:ODE}. 

\begin{figure}
    \centering
    \includegraphics[width=1\linewidth]{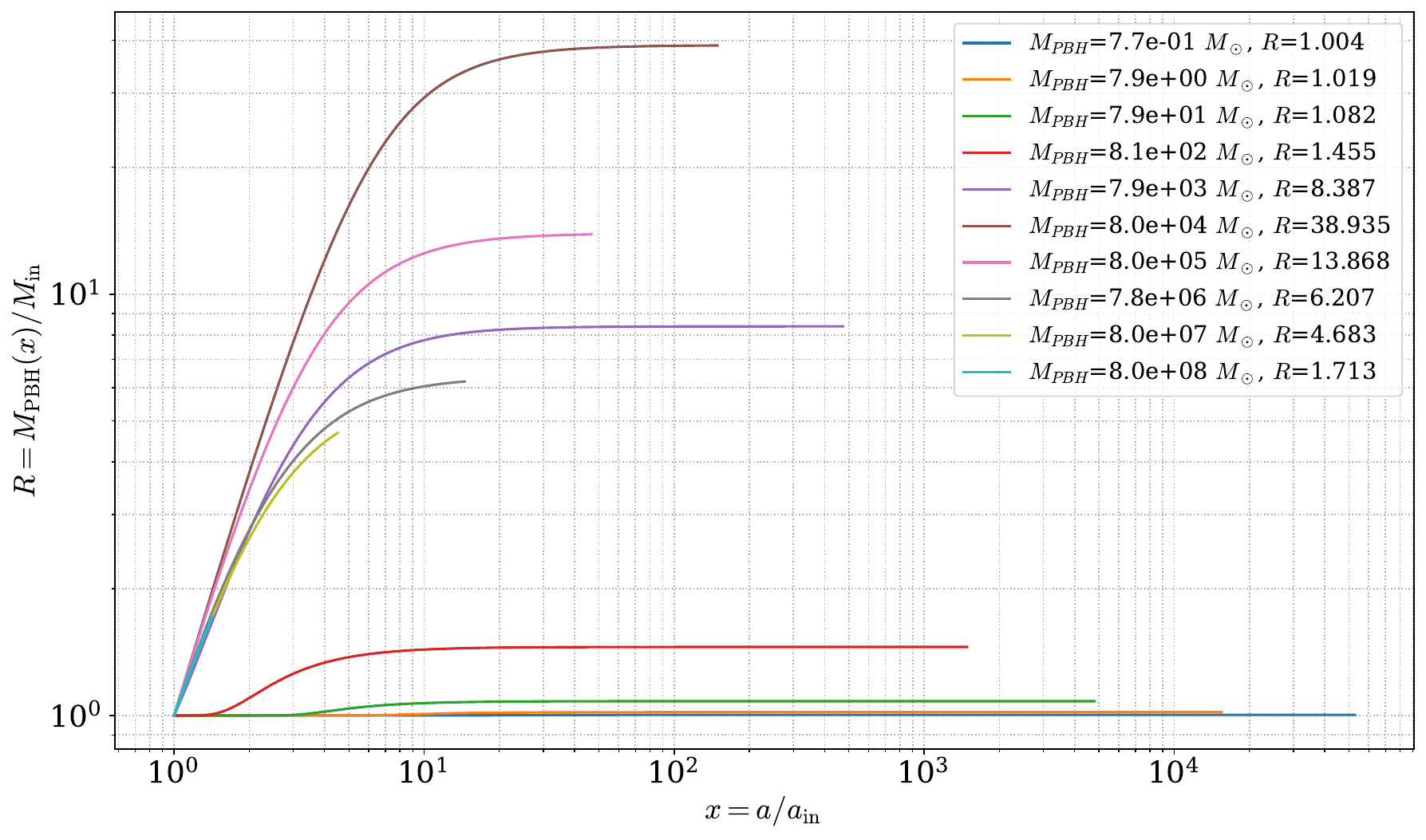}
    \caption{Mass growth $R$ against $x$, the measure of time after formation with $\gamma=0.55$ and $\kappa=1$.}
    \label{fig:Mass_growth_bis}
\end{figure}

The saturation of the mass growth can be seen in Fig.~\ref{fig:Mass_growth_bis}, where we plot the time evolution of various PBH masses. We see that the mass growth $R$ does not depend linearly on the PBH mass; the relevant quantities to understand the mass growth hierarchy are once again $x$ and $\delta_{hf}/\alpha \propto g_\rho^{\nu}/g_\rho^{\rm tot}$. The relative contribution of neutrinos peaks between $T=10 \rm ~MeV$ and $1\rm~MeV$, following the QCD transition. The PBHs formed in this window with $M\sim10^3-10^5M_\odot$ have a lot of radiation available to absorb, and their $T_{\rm init}\sim T_{\rm start}$. This temperature window is the sweet spot for neutrino absorption, and we see $R$ peaking in this region.

At smaller temperatures $T\sim1 \rm~MeV$, neutrinos are decoupling from the plasma, and $g_\rho^{\nu}/g_\rho^{\rm tot}$ decreases before freezing until the end of radiation at $T_{\rm eq}$. The associated heavy PBHs, even though they can absorb from their formation to $T_{\rm eq}$, do not grow has much as the intermediate mass window because neutrinos are a subdominant source of radiation. We see this with $M_{\rm PBH}\sim10^6 M_\odot$. Moreover, the later the PBH is born, the smaller the absorption window is $[T_{\rm init};~T_{\rm eq}]$; heavy PBHs have $T_{\rm start}=T_{\rm init}$. Hence, PBHs with $M_{\rm PBH}\gtrsim10^6 M_\odot$ do not have time to reach saturation.

\begin{figure}
    \centering
    \includegraphics[width=\linewidth]{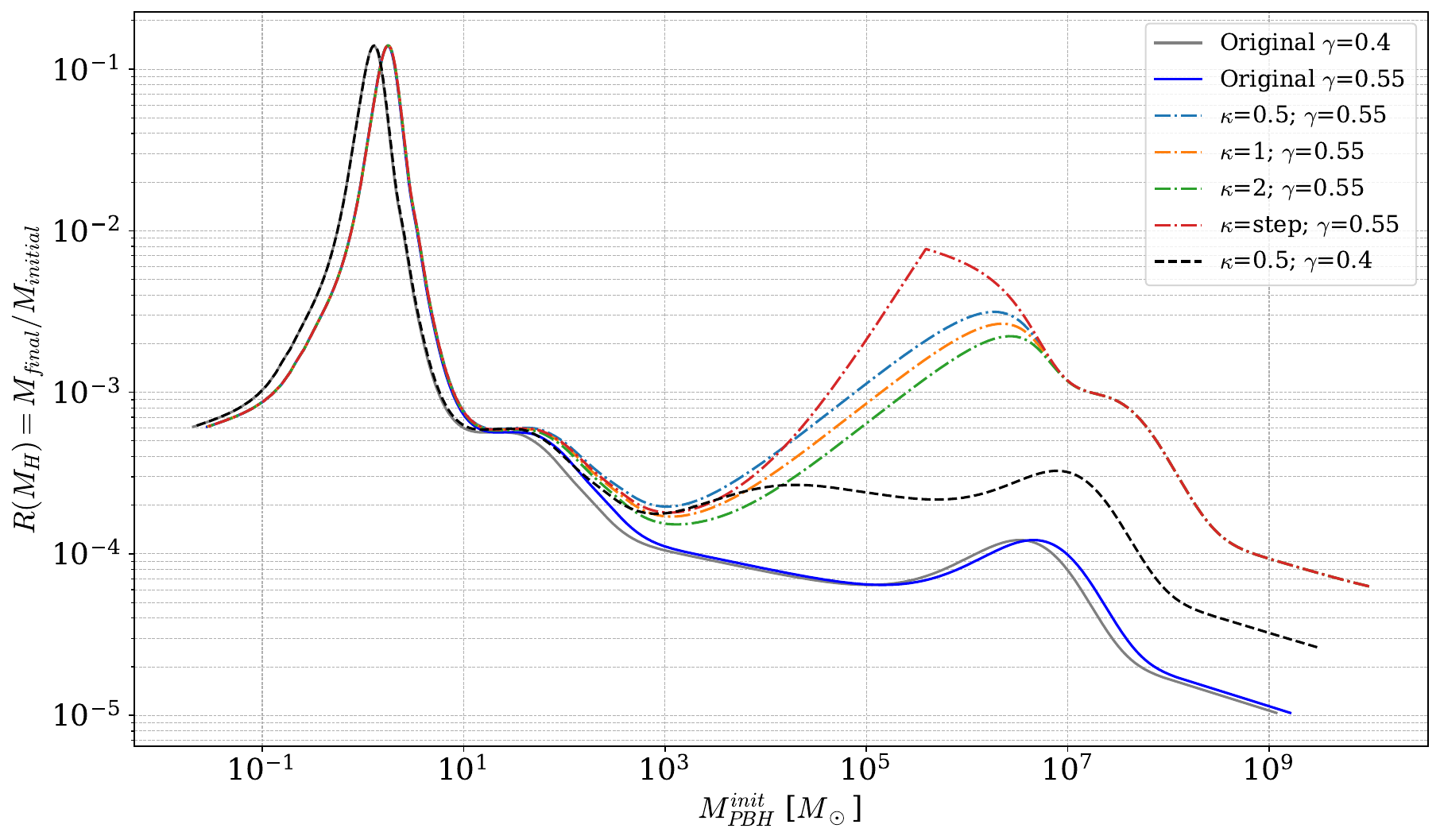}
    \caption{PBH mass spectra initially normalized at $f_{\rm PBH}^{\rm rad}=0.1$ (solid line 'Original' in the plot), with $\gamma=0.55$ for dash-dotted lines and $\gamma=0.4$ for the black dashed line. Naturally reducing the value of $\gamma$ shifts the QCD peak to lower values. To clarify the overlapping curves: the QCD peak and surrounding regions overlap for models sharing the same value of $\gamma$.}
    \label{fig:PBH_spectra}
\end{figure}

With all the effects mentioned above in mind, we now plot the updated PBH mass spectra in Fig.~\ref{fig:PBH_spectra} from Eq.~\eqref{eq:abs_spectrum}. Neutrino absorption leaves distinct imprints in the spectrum: it increases the contribution of post-QCD transition PBHs with $M_{\rm PBH}\gtrsim10^2 M_\odot$, shifts the $e^+e^-$ peak to larger masses, modifies the value of $f_{\rm PBH}$, and creates an additional peak. The first three consequences are easily understood: heavy PBHs gain mass, which increases their contribution, shifts the corresponding structures to larger masses, and as a result, the fraction of dark matter in PBHs increases $f_{\rm PBH}$. Note that the updated value $f_{\rm PBH}^{\rm abs}$ depends on the absorption transition; the step function yields the highest value of $f_{\rm PBH}^{\rm abs}$, the sharper the transition, the larger $f_{\rm PBH}^{\rm abs}$, see Table~\ref{tab:kappa_fPBH}. 

\begin{table}[h]
\centering
\begin{tabular}{c|cccc}
\hline
$\kappa$ & Step & 0.5 & 1 & 2 \\
\hline
$f_{\rm PBH}^{\rm abs}$ & 0.1260 & 0.1156 & 0.1129 & 0.1107 \\
\end{tabular}
\caption{DM in PBH fraction as a function of $\kappa$ parameter for $\gamma=0.55$ and $f_{\rm PBH}^{\rm rad}=0.1$}
\label{tab:kappa_fPBH}
\end{table}

About the additional peak, it is this time relative to the absorption sweet spot and $\gamma$. PBHs born in the right conditions, mentioned in the paragraph above, will gain more mass compared to the rest of the spectrum and accumulate in an additional peak. Comparing Fig.~\ref{fig:Mass_growth} and Fig.~\ref{fig:PBH_spectra}, we see the same structure from $R(M_{\rm PBH})$  appearing in the PBH mass spectra. Since $R$ depends on $\gamma$, the shifts of existing structures, $f_{\rm PBH}^{\rm abs}$ (see Table\ref{tab:gamma_fPBH}), the additional peak amplitude, and position also depend on $\gamma$. 

\begin{table}[h]
\centering
\begin{tabular}{c|ccccc}
\hline
$\gamma$ & 0.1 & 0.3 & 0.4 & 0.55  \\
\hline
$f_{\rm PBH}^{\rm abs}$ & 0.1006 & 0.1019 & 0.1031 & 0.1129 \\
\end{tabular}
\caption{DM in PBH fraction as a function of $\gamma$ parameter for $\kappa=1$ and $f_{\rm PBH}^{\rm rad}=0.1$}
\label{tab:gamma_fPBH}
\end{table}

Moreover, we can apply upper bounds on $\gamma$ in two ways: if $f_{\rm PBH}^{\rm abs}>1$ or if the PBHs undergo a runaway absorption. The latter is the most stringent for $f_{\rm PBH}^{\rm rad}=0.1$ in the mass range considered (see Appendix \ref{app:MassGrowth}). From Eq.~\eqref{eq:gamma_div}: $\gamma_{div}>0.55$. Above this value, the neutrino absorption of PBHs in the temperature sweet spot runs away and mass growth diverges. Note that the bound $\gamma\lesssim0.55$ apply on $\gamma$ and not $\gamma_{\rm eff}$. 
These bounds must be taken with a grain of salt: as noted by Barrau et al. \cite{Barrau:2022bfg,Barrau:2025psm}, the runaway absorption may be an artifact of the Schwarzschild description pushed beyond its domain of validity, and a regularized metric could soften the divergence — in which case our bounds on $\gamma$ are conservative. 

We now briefly discuss the consequences of neutrino absorption on PBH observations and 
constraints. As the aim of this letter is to highlight the mechanism, we restrict ourselves 
to a qualitative discussion. As in the late-time accretion scenario of De Luca et 
al.~\cite{DeLuca:2020bjf,DeLuca:2020fpg}, constraints must be remapped from formation 
mass to observed mass. Although the mechanism presented here still occurs during the 
radiation era, the CMB and accretion bounds~\cite{Serpico:2020ehh,Facchinetti:2022kbg,
Agius:2024ecw} are not relieved by absorption; if anything, growth into this mass window 
tightens them, as they apply to the final mass and abundance. The QCD peak at 
$M_{\rm PBH} \sim 1\,M_\odot$ and the lower portion of the spectrum are unaffected, and 
the current debate on the constraints remains unchanged; see~\cite{Franciolini:2022tfm,
LISACosmologyWorkingGroup:2023njw,Carr:2023tpt,Carr:2026hot}.

Since mass growth only significantly impacts $M_{\rm PBH} \gtrsim 10^3\,M_\odot$ 
(see Figs.~\ref{fig:Mass_growth} and~\ref{fig:Mass_growth_bis}), it has implications for 
structure formation~\cite{Carr:2018rid}, intermediate-mass BH observations~\cite{
2024Natur.631..285H,Huang:2024gpv}, and James Webb Space Telescope `little red dots' 
(LRDs); see Refs.~\cite{Bogdan:2023ilu,2024Natur.627...59M,Matthee:2023utn,
2023ApJ...957L...3P} for observations and Refs.~\cite{Liu:2022bvr,Dayal:2024zwq,Zhang:2025oyl,
DeLuca:2025nao} for discussion of LRDs and PBHs.

One striking takeaway from our study is the disentangling of large density fluctuations from intermediate-mass and supermassive PBHs. Through neutrino absorption, the Universe can form heavy PBHs, $M_{\rm PBH} \gtrsim 10^4M_\odot$ without requiring large density fluctuations, thereby partially evading the $\mu-$~distortion \cite{Chluba:2012we,Nakama:2017xvq,Byrnes:2024vjt} and scalar-induced gravitational-wave bounds \cite{Domenech:2021ztg,2024MNRAS.528..883C,Cecchini:2025oks}, which constrain the fluctuation at the formation scale rather than the final mass, independently of the non-Gaussianity evasion route \cite{Nakama:2017xvq,Byrnes:2024vjt}.  Additionally, in the presence of large extra dimensions, microscopic PBHs could undergo mass growth by several orders of magnitude~\cite{Vitale:2026ric}.

\section{Summary and Conclusions}
\label{sec:summary}
In this study, we showed that significant PBH mass growth is possible in the radiation era for $M_{\rm PBH}\sim10^3-10^7M_\odot$ with $\gamma=0.55$. Using a semi-classical description of a BH in a thermal bath and a closer look at the kinetics of the primordial plasma, we find that neutrinos can be absorbed by PBHs. Depending on the collapse fraction $\gamma$, the resulting mass growth can create additional features in the PBH mass spectrum and increase the portion of dark matter in PBHs. We examined four models of the transition from non-absorbing to absorbing PBHs, finding graphical differences but small variations in $f_{\rm PBH}^{\rm abs}$.

In this letter, we coupled leptons and PBHs, but we did not account for lepton or baryon asymmetries of the Universe. We defer such work to future studies. 
These results have implications for dark matter scenarios involving PBHs. The absorption sweet spot can induce an additional peak in the PBH mass distribution between $M_{\rm PBH}\sim10^3-10^7M_\odot$, although the effect is more significant for more massive ones. 

PBHs are already candidates for the James Webb Space Telescope's s; neutrino-absorbing PBHs could be the culprit. As light seeds can grow rapidly, these PBHs evade constraints based on primordial fluctuations. Thanks to neutrino absorption, the production of intermediate-mass PBHs with $M_{\rm PBH}\sim10^3-10^5M_\odot$ does not rely solely on sizable fluctuations. The PBH mass spectrum is known to be a probe of the early Universe; we have pointed out a mechanism distorting it and revised the claim that PBHs cannot grow in mass in the radiation era.

\begin{acknowledgments}
I would like to thank my PhD supervisors G\"unther Hasinger and David Blaschke for their 
guidance and the freedom they have allowed me to enjoy. I also thank Florian K\"uhnel, 
David Kaiser, Albert Escriv\`a, Oleksii Ivanytskyi  and Julien Froustey for their time, 
and Kazunori Kohri, Dejan Stojkovic, and Shibsankar Si for their interest in the first 
version of the manuscript published on arXiv.
The author gratefully acknowledge the financial support provided by the German Federal Ministry of Research, Technology and Space (BMFTR) in the framework of the Knowledge creates perspectives for the region!, for the project StStG – DZA – Aufbauphase: Deutsches Zentrum für Astrophysik, Großforschungszentrum in der sächsischen Lausitz: Aufbauphase 2026, grant number 03WSP1746.
\end{acknowledgments}

\appendix

\section{PBH mass spectrum}\label{app:pbh_spectrum}
If $\delta > \delta_c$ at $t_{\rm cross}$ a PBH is formed, if not the overdensity is eventually dispersed away by the pressure.
Assuming Gaussian fluctuations, the fraction of the Universe collapsing is
\begin{equation}\label{eq:beta}
    \beta(M)\approx \mathrm{erfc}\left[\frac{\delta_c(w(T(M))}{\sqrt{2}\delta_{\rm rms}(M)}\right]~,
\end{equation}
where $M$ is the PBH mass, $\rm erfc$ is the complementary error function and $\delta_{\rm rms}$ is the root mean square amplitude of the Gaussian fluctuation; $\delta_c(w(T(M)))$ is taken from \cite{Musco:2012au}. Following \cite{Bodeker:2020stj,Carr:2019kxo},
\begin{equation}\label{eq:delta_rms}
    \delta_{\rm rms} = A\times(M/M_\odot)^{(1-n_s)/4},
\end{equation}
where $n_s=0.97$ is the spectral index taken at its CMB value \cite{Planck:2018vyg}. On PBH scales there is actually more liberty on the shape of $\delta_{rms}$ and the value of $n_s$, for instance $n_s$ could be running \cite{Braglia:2021wwa}. The amplitude $A$ is a normalization parameter that expresses the strength of the fluctuations.

The present fraction of DM in a PBH of mass $M$ is then
\begin{equation}
\frac{df_{\rm PBH}(M)}{d \ln M} \approx 2.4 ~\beta(M)\sqrt{M_{\rm eq}/M}~,
\end{equation}
where $M_{\rm eq}$ is the horizon mass at matter-radiation equality. The numerical factor originates from
$ 2.4=2(1+\Omega_b/\Omega_{\rm CDM})$, with $\Omega_b=0.0456$ and $\Omega_{\rm CDM}=0.245$ being the baryon and CDM density parameters from \cite{Planck:2018vyg}. 

\begin{equation}\label{eq:f_pbh}
    f_{\rm PBH} \equiv \int_{M_{\rm min}}^{M_{\rm max}} \frac{df_{\rm PBH}}{d \ln M} \,d \ln M\,.
\end{equation}

\section{Details on the Mass Growth Calculation}
\label{app:MassGrowth}
In this appendix we rewrite the equations from Ref.~\cite{Haque:2026vvp}, define the absorption regime and discuss the introduction of $\gamma_{\rm eff}$.

\subsection{Geometrical cross section}

A BH in a radiation bath has an associated cross section, the formula giving it depends on the absorption regime. The high frequency $\omega  \gg 1/r_S$ and low frequency regime $\omega \ll 1/r_S$. In our mass range, the high frequency regime, also written with $r_ST\gg1$ is fulfilled at all time. The PBH geometrical absorption cross section follows \cite{1966SvPhU...8..522Z,Zeldovich:1967lct,Custodio:1998pv,Custodio:2002gj}:
\begin{equation}
    \sigma_{hf} = \frac{27}{64\pi} \frac{M_{\rm BH}^2}{M_{Pl}^4} = \frac{27}{64\pi} \frac{1}{T_{\rm BH}^2}
\end{equation}
In our mass range the BH cross section is huge. Although we consider absorption from neutrinos only, we expect some significant mass growth.
The mass evolution is given by:

\begin{equation}\label{eq:absRate}
    \frac{dM_{\rm BH}}{dt}=\sigma_{hf}\rho_R=\delta_{hf}\frac{T^4}{T_{\rm BH}^2}
\end{equation}

where $\rho_R$ the radiation energy density, and $g_\rho=\frac{30\rho}{\pi^2T^4}$ the relativistic degrees of freedom.
\begin{equation}\label{eq:delta_hf}
    \delta_{hf}=\frac{9\pi}{640}g_\rho
\end{equation}

Note that in the geometric-optics limit $\sigma_{hf}$ is spin-blind; the dependence on the nature of the radiation enters only through $\delta_{hf} \propto g_\rho$, i.e., the fermionic statistics of neutrinos affect the absorbed energy density, not the cross section.
Only neutrinos are absorbed in our framework, and $\delta_{hf}$ is relative to the PBH-radiation cross section therefore $g_\rho\equiv g_\rho^\nu$ in Eq.~\eqref{eq:delta_hf}. On the other hand $\alpha$ is relative to the cosmic expansion (see Eq.~2 in Ref.~\cite{Haque:2026vvp}), so that $\alpha=g^{total}_\rho \pi^2/30$.
Since we neglect BH evaporation Eq.~\eqref{eq:ODE} can be written as follow. We start by defining the function $g(x)=\frac{12\pi\gamma_{\rm eff}\delta_{hf}}{\alpha}x^3 W$ and $G(x)=\int dxg(x)$ to write the ODE as:
\begin{equation}
    \frac{dR}{dx}=g(x)R^2
\end{equation}
It follows that :
\begin{equation}
    R = \frac{1}{1-G(x)}
\end{equation}
 
Although the dependency of $G$ on $x$ seems obvious at first sight, in our definition of $g(x)\propto \rho_\nu/\rho_{total}$, a ratio depending on the scale factor as well. $\gamma_{\rm eff}\propto(1+\delta_c(T))$ is also a function of the scale factor. Therefore $G(x,\gamma)$ is determined numerically. 
The general condition for runaway absorption is $G(x,\gamma)\geqslant1$. But we can also solve analytically the integral $G(x)=\int dxg(x)$ as long as the temperature range considered have $\rho_\nu/\rho_{total}=\rm const$ and $\delta_c=\rm const$. Conveniently, on the absorption sweet spot both condition are fulfilled, therefore we can use the bound from Eq.~21 of Ref.~\cite{Haque:2026vvp} and get to Eq.~\eqref{eq:gamma_div}.

\subsection{On $\gamma_{\rm eff}$}
In the main text we defined an effective collapse fraction $\gamma_{\rm eff}\propto(1+\delta_c)$ defined by the overdensity, but $a_{\rm rad}$ is defined on the background. $\gamma_{\rm eff}$ is now a function of the temperature as well. During phase transition $\delta_c$ value dip hence $\gamma_{\rm eff}$ follows. Because the universe gets softer, the energy overdensity threshold is smaller and the PBH, if formed, have smaller masses compared to background hubble mass.

On the other hand, without the factor $(1+\delta_c)$ the resulting PBH mass is artificially weighted down to smaller values. Even with a $\gamma=1$ the PBH mass at formation will correspond to any background hubble volume, missing the excess that define the overdensity able to form a PBH.

\nocite{*}

\bibliography{apssamp}

\end{document}